\documentclass{emulateapj}
\usepackage{apjfonts}
\usepackage{color}

\newcommand{\pivec}{\mbox{\boldmath $\pi$}}

\lefthead{HAN ET AL.} 
\righthead{SPACE-BASED MICROLENS PARALLAX OBSERVATION}

\begin{document}

\title{Space-based Microlens Parallax Observation As a Way to Resolve the Severe Degeneracy between 
Microlens-parallax and Lens-orbital Effects}

\author{
C.~Han$^{1}$, A.~Udalski$^{O1,O}$, C.-U.~Lee$^{K1,K}$, A.~Gould$^{U1,U2,K1,K}$, V.~Bozza$^{I1,I2}$ \\
and \\
M.~K.~Szyma{\'n}ski$^{O1}$, I.~Soszy{\'n}ski$^{O1}$, J.~Skowron$^{O1}$, P.~Mr{\'o}z$^{O1}$, 
R.~Poleski$^{O1,U1}$, P.~Pietrukowicz$^{O1}$, S.~Koz{\l}owski$^{O1}$, K.~Ulaczyk$^{O1}$, 
{\L}.~Wyrzykowski$^{2}$, M.~Pawlak$^{O1}$\\
(The OGLE Collaboration),\\
M.~D.~Albrow$^{K2}$, S.-J. Chung$^{K1}$, S.-L.~Kim$^{K1}$, S.-M.~Cha$^{K1,K3}$,
Y.~K.~Jung$^{K4}$, D.-J. Kim$^{K1}$, Y.~Lee$^{K1,K3}$, B.-G.~Park$^{K1}$, Y.-H. Ryu$^{K1}$, 
I.-G.~Shin$^{K4}$, J.~C.~Yee$^{K4,F1}$ \\
(The KMTNet Collaboration),\\
}

\affil{$^{1}$  Department of Physics, Chungbuk National University, Cheongju 361-763, Republic of Korea}
\affil{$^{O1}$ Warsaw University Observatory, Al. Ujazdowskie 4, 00-478 Warszawa, Poland}
\affil{$^{U1}$ Department of Astronomy, Ohio State University, 140 W. 18th Ave., Columbus, OH 43210, USA}
\affil{$^{U2}$ Max Planck Institute for Astronomy, K{\"o}nigstuhl 17, D-69117 Heidelberg, Germany}
\affil{$^{I1}$ Dipartimento di Fisica "E.~R.~Caianiello", U\'niversit\'a di Salerno, Via Giovanni Paolo II, I-84084 Fisciano (SA), Italy}
\affil{$^{I2}$ Istituto Internazionale per gli Alti Studi Scientifici (IIASS), Via G. Pellegrino 19, I-84019 Vietri Sul Mare (SA), Italy}
\affil{$^{K1}$ Korea Astronomy and Space Science Institute, Daejeon 34055, Republic of Korea}
\affil{$^{K2}$ University of Canterbury, Department of Physics and Astronomy, Private Bag 4800, Christchurch 8020, New Zealand}
\affil{$^{K3}$ School of Space Research, Kyung Hee University, Yongin 17104, Republic of Korea}
\affil{$^{K4}$ Harvard-Smithsonian Center for Astrophysics, 60 Garden St., Cambridge, MA, 02138}
\footnotetext[O]{The OGLE Collaboration.}
\footnotetext[K]{The KMTNet Collaboration.}
\footnotetext[F1]{Sagan Fellow.}

\begin{abstract}
In this paper, we demonstrate the severity of the degeneracy between the 
microlens-parallax and lens-orbital effects by presenting the analysis of 
the gravitational binary-lens event OGLE-2015-BLG-0768. Despite the obvious 
deviation from the model based on the the linear observer motion and the 
static binary, it is found that the residual can be almost equally well 
explained by either the parallactic motion of the Earth or the rotation 
of the binary lens axis, resulting in the severe degeneracy between the 
two effects.  We show that the degeneracy can be readily resolved with 
the additional data provided by space-based microlens parallax observations.  
Enabling to distinguish between the two higher-order effects, space-based 
microlens parallax observations will make it possible not only to accurately 
determine the physical lens parameters but also to further constrain the 
orbital parameters of binary lenses.
\end{abstract}


\section{INTRODUCTION}

\color{black}

Objects that are bound together by gravity move along orbits following Kepler's law. 
In gravitational microlensing, the orbital motion of the Earth around the Sun causes the 
motion of an observer to deviate from rectilinear. The modulation of the observer's motion 
is reflected onto the relative lens-source position due to parallax effects, resulting in 
deviations in lensing light curves from those expected from a rectilinear motion of the 
observer \citep[microlens parallax effect:][]{Gould1992}. For lenses composed of two masses, 
on the other hand, the orbital motion of the binary lens induces the rotation of the binary 
axis. This also causes modulations of the relative lens-source position and deviations from 
the light curve of a static binary-lens event 
\citep[lens orbital effect:][]{Dominik1998, Ioka1999}.

Detecting deviations in lensing light curves caused by the orbital motions of the observer 
and the lens are important because they can provide us with a useful information that can 
be used to characterize lens systems. Analysis of lensing light curves affected by the 
microlens parallax effect enables one to measure the microlens parallax vector $\pivec_{\rm E}$, 
of which magnitude is related to the physical parameters of the lens mass $M$ and the distance 
to the lens $D_{\rm L}$ by 
\begin{equation}
M={\theta_{\rm E}\over \kappa \pi_{\rm E}};\qquad
D_{\rm L}={{\rm AU}\over \pi_{\rm E}\theta_{\rm E}+\pi_{\rm S}},
\end{equation}
where $\theta_{\rm E}$ is the angular Einstein radius, $\kappa=4G/(c^2{\rm AU})$,
AU is the astronomical unit, $\pi_{\rm S}={\rm AU}/D_{\rm S}$, and $D_{\rm S}$ is the distance 
to the source \citep{Gould2000}. Analyses of the light curves affected by the lens-orbital 
effect enables one to constrain the orbital parameters of a binary lens system \citep{Shin2011}.

Rooted on the same origin of the orbital motion, however, both the parallactic motion of the 
Earth and the rotation of the binary lens may have similar effects on the relative lens-source 
motion, resulting in similar deviations in lensing light curves \citep{Batista2011, Skowron2011}. 
If so, characterizing binary lenses by detecting the microlens-parallax and lens-orbital effects 
can be seriously hampered due to the difficulty in distinguishing one effect from the other.

In addition to the single frame of the accelerating Earth, microlens parallaxes can also be 
measured from the simultaneous observation of a lensing event using ground-based telescopes 
and a space-based satellite in a solar orbit. In this case, the projected Earth-satellite 
separation $D_\perp$ is comparable to the Einstein radius of typical Galactic microlensing 
events, i.e.\ $\sim (O)$ AU, and thus the relative lens-source positions seen from ground 
and in space appear to be different, resulting in different light curves. Combined analysis 
of the light curves obtained from the ground-based and space-based observations leads to the 
measurement of the microlens parallax \citep{Refsdal1966, Gould1994}, which is referred to 
as the ``space-based microlens parallax''. In order to distinguish from the space-based 
microlens parallax, the microlens parallax measured based on the annular parallactic motion 
of the Earth is referred to as ``annual microlens parallax''.

The space-based microlens parallax measurement was recently realized by the microlensing 
program using the {\it Spitzer Space Telescope} \citep{Gould2014}. The goal of the program 
is to determine the Galactic distribution of planets by measuring microlens parallaxes and 
thereby estimate the distances of the individual lenses \citep{Calchi2015}. From the 
observations conducted in 2014 and 2015 seasons, the {\it Spitzer} microlensing program yielded 
important scientific results including measurements of the physical parameters of two 
planetary systems \citep{Udalski2015, Street2016}, microlens parallax measurements of 22 
single-mass objects \citep{Yee2015, Calchi2015}, characterizations of binary objects including 
the discovery of a binary with a massive remnant component \citep{Zhu2015, Shvartzvald2015, 
Bozza2016, Han2016}. Followed by the successful first two seasons, space-based microlensing 
observations using the {\it Spitzer} mission will be carried on in 2016 season. In addition to 
the {\it Spitzer} microlensing program, {\it K2}'s Campaign 9 is scheduled to conduct a microlensing 
survey in 2016 season. Since {\it K2} has an Earth-trailing heliocentric orbit with a semi-major 
axis $\sim 1.0$ AU, it will be also an important instrument for space-based microlens 
parallax observation. From the {\it K2} survey, it is expected to measure microlens parallax 
for $\gtrsim 127$ lensing events \citep{Henderson2016}.

Space-based microlens parallax observations can be useful in resolving the possible 
degeneracy between the microlens-parallax and lens-orbital effects in the analyses of 
lensing light curves. The positional change of an observer during a lensing event caused 
by the annual parallactic motion of the Earth is usually small and thus deviations in 
lensing light curves are subtle. In contrast, the difference between the light curves 
seen from ground and in space is very prominent because the Earth-satellite separation 
is a significant portion of the Einstein radius. Since the difference is almost entirely 
attributed to the parallax effect, then, the microlens parallax can be uniquely determined 
from the combined analysis of the two light curves. Once the microlens parallax is measured, 
one can constrain the orbital lens parameters by further analyzing the residual from the 
model with parallax parameters.

In this paper, we demonstrate the severity of the degeneracy between the microlens-parallax 
and lens-orbital effects by presenting the analysis of an actually observed binary-lens event. 
We also show that the degeneracy can be readily resolved with the additional data from 
space-based parallax observations.

\begin{figure}[t]
\epsscale{1.2}
\plotone{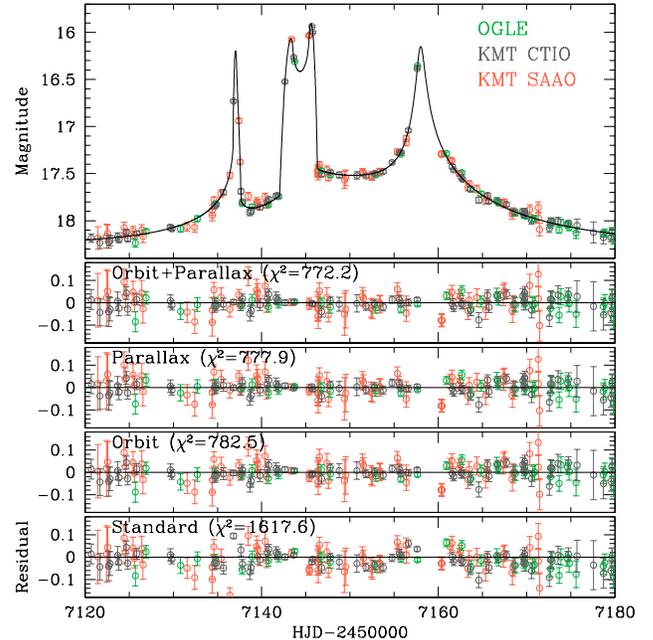}
\caption{\label{fig:one}
Light curve of OGLE-2015-BLG-0768. Lower panels show residuals from various tested 
models and the $\chi^2$ values of the models.  The curve superposed on the observed 
data is the best-fit model based on the ``orbit+parallax'' model.
}
\end{figure}

\section{Degeneracy: OGLE-2015-BLG-0768 Case} 

In order to describe a binary lensing light curves, one needs many parameters. A microlensing 
modeling is a process of finding a set of lensing parameters that results in the best-fit model 
light curve to the observed one. For the simplest case of a binary-lens event with a linear 
observer's motion and a static binary, one needs 7 parameters and the number 
of parameters increases in order to consider higher-order effects like the microlens-parallax 
and lens-orbital effects. Furthermore, lensing parameters are usually tightly correlated one 
another \citep{Han2010} and thus the parameters describing the microlens-parallax and lens-orbital 
effects are likely to be related not only to each other but also to other parameters. As a result, 
it is difficult to analytically track down the complex multi-dimensional correlations between 
parameters. We, therefore, show the severity of the degeneracy between the two higher-order 
effects for an example event where higher-order effects are needed to describe the observed 
light curve.

OGLE-2015-BLG-0768 is an exemplary lensing event where we find that the degeneracy between the 
microlens-parallax and lens-orbital effects is severe. The event occurred on a source star that 
is located toward the Galactic bulge field. The coordinates of the source are 
$({\rm RA},{\rm DEC})_{\rm J2000} =(17^{\rm h}38^{\rm m}19^{\rm s}\hskip-2pt.15, 
-27^\circ 28'02''\hskip-2pt.5)$, that correspond to 
the Galactic coordinates $(l,b)=(0.40^\circ,2.14^\circ)$. The lensing-induced brightening of the 
source star was discovered on 2015 April 22  by the Early Warning System (EWS) 
of the Optical Gravitational Lensing Experiment (OGLE) group that has conducted microlensing survey 
since 1992 using the 1.3m telescope located at the Las Campanas Observatory in Chile \citep{Udalski1992}. 
The event was also observed by the Korean Microlensing Telescope Network \citep[KMTNet:][]{Kim2016} 
survey that was commenced in 2015. The KMTNet survey is composed of three identical 1.6m telescopes 
that are located at Cerro Tololo Interamerican Observatory in Chile (KMT CTIO), South African 
Astronomical Observatory in South Africa (KMT SAAO), and Siding Spring Observatory in Australia 
(KMT SSO). The KMT CTIO and SAAO started their operation in February, 2015, but the KMT SSO was 
not yet operational at the time of the event. The KMTNet survey conducted with a $\sim 10$ minute 
cadence for its main field. The source star of the event was in the field that was observed 
with a $\sim 1/2$ -- 1 day cadence mainly in support of the {\it Spitzer} microlensing program.

In Figure~\ref{fig:one}, we present the light curve of OGLE-2015-BLG-0768. It shows a complicated 
sequence of three peaks that occurred chronologically at ${\rm HJD}'={\rm HJD}-2450000\sim 7137$, 
7144, and 7158. The second peak is composed of two spikes where the region between them shows a 
``U''-shape trough. Such an anomaly is a characteristic feature that appears when a source crosses 
caustics, which represent the locations on the source plane at which a point-source magnification 
becomes infinity, formed by a binary lens, indicating that the event was occurred by a binary lens. 
The first and third peaks do not show such a caustic-crossing feature, suggesting that they were 
produced by either passing over or approaching tips of the caustic.

Considering the characteristic features of the light curve, we conduct binary-lens modeling 
of the light curve. We begin with a simplest case where the relative lens-source motion is 
linear. Among the 7 principal binary-lensing parameters for this standard modeling, 3 parameters 
describe the source-lens approach, including the time of the closest approach of the source to 
a reference position of the lens, $t_0$, the source-reference separation at that time, $u_0$, 
and the angle between the source trajectory and the binary axis, $\alpha$ (source trajectory 
angle).  For the reference position in the lens plane, we use the center of mass of the binary 
lens.  Another parameter $t_{\rm E}$ (Einstein time scale), which is defined as the time for 
the source to cross the angular Einstein radius $\theta_{\rm E}$ of the lens, represents the 
time scale of the event. The Einstein ring is the image of the source for the case of the exact 
lens-source alignment and it is used as a length scale in describing lensing phenomenon.  Two 
other parameters characterize the binary lens including $s$ and $q$, which represent the projected 
separation and mass ratio, respectively.  We note that the two parameters $u_0$ and $\alpha$ are 
normalized to $\theta_{\rm E}$.  The last parameter $\rho$ is the ratio of the angular source 
radius $\theta_*$ to the angular Einstein radius, i.e. $\rho=\theta_*/\theta_{\rm E}$. This 
normalized source radius is needed to describe the caustic-crossing parts of the light curve 
that are affected by the finite size of the source star.

\begin{deluxetable*}{lrrrr}
\tablecaption{Lensing parameters\label{table:one}}
\tablewidth{0pt}
\tablehead{
\multicolumn{1}{c}{Parameters}      &
\multicolumn{1}{c}{Standard}        &
\multicolumn{1}{c}{Parallax}        &
\multicolumn{1}{c}{Orbit}           &
\multicolumn{1}{c}{Orbit+Parallax} 
}
\startdata
$\chi^2$                       & 1617.6                & 777.9                & 782.5                & 772.2                 \\     
$t_0 \ ({\rm HJD}-2450000)$    & 7156.157 $\pm$ 0.049  & 7154.669 $\pm$ 0.152 & 7155.099 $\pm$ 0.074 & 7154.931 $\pm$ 0.112  \\     
$u_0$                          & 0.214    $\pm$ 0.001  & 0.253    $\pm$ 0.001 & 0.249    $\pm$ 0.001 & 0.248    $\pm$ 0.003  \\     
$t_{\rm E} \ ({\rm days})$     & 19.69    $\pm$ 0.04   & 20.35    $\pm$ 0.16  & 19.39    $\pm$ 0.06  & 20.08    $\pm$ 0.21   \\     
$s$                            & 0.773    $\pm$ 0.001  & 0.778    $\pm$ 0.001 & 0.768    $\pm$ 0.002 & 0.763    $\pm$ 0.006  \\     
$q$                            & 0.849    $\pm$ 0.019  & 0.791    $\pm$ 0.024 & 0.858    $\pm$ 0.018 & 0.798    $\pm$ 0.023  \\     
$\alpha \ ({\rm rad})$         & 1.179    $\pm$ 0.008  & 0.986    $\pm$ 0.015 & 1.023    $\pm$ 0.009 & 1.008    $\pm$ 0.011  \\     
$\rho \ (10^{-3})$             & 7.76     $\pm$ 0.23   & 8.40     $\pm$ 0.19  & 7.76     $\pm$ 0.23  & 8.08     $\pm$ 0.25   \\ 
$\pi_{{\rm E},N}$              & -                     & -4.52    $\pm$ 0.30  & -                    & -2.26    $\pm$ 1.00   \\              
$\pi_{{\rm E},E}$              & -                     & 0.12     $\pm$ 0.25  & -                    & -0.08    $\pm$ 0.22   \\              
$ds/dt$ $(yr^{-1})$            & -                     & -                    & -0.23    $\pm$ 0.06  & -0.46    $\pm$ 0.18   \\
$d\alpha/dt$ $(yr^{-1})$       & -                     & -                    & -2.55    $\pm$ 0.05  & -0.89    $\pm$ 0.57   
\enddata
\end{deluxetable*}

Modeling the light curve is proceeded in multiple steps. In the first step, we conduct a grid 
search in the space of the parameters $s$, $q$, and $\alpha$ for which lensing light curves vary 
sensitively to the changes of the parameters. We search for the other parameters by using a downhill 
approach based on the Markow Chain Monte Carlo (MCMC) method. In the second step, we identify 
local minima in the $\chi^2$ map of the parameters. We identify a unique minimum with a projected 
binary separation $s\sim 0.8$ and a mass ratio $q\sim 0.8$. In the next step, we gradually 
refine the identified minimum by allowing all parameters to vary.

We compute magnifications affected by the finite size of the source star using the 
combination of the numerical inverse-ray-shooting method \citep{Schneider1986} and 
the semi-analytic hexadecapole approximation \citep{Pejcha2009, Gould2008}. In computing 
finite-source magnifications, we consider the surface-brightness variation of the source 
star caused by the limb-darkening effect by modeling the surface brightness profile as 
$S_\lambda \propto 1-\Gamma_\lambda (1-3\cos\psi/2)$, where $\psi$ is the angle between 
the line of sight toward the source center and the normal to the source surface and 
$\Gamma_\lambda$ is the linear limb-darkening coefficient. From the measurements of 
$I$ and $V$-band magnitudes of the source star followed by the calibration of the 
color and magnitudes using the centroid of giant clump in the color-magnitude diagram 
as a reference \citep{Yoo2004}, we find that the source star is a G-type giant with a 
de-reddened color and a magnitude $(V-I, I)_0=(0.86\pm 0.02,15.92\pm 0.01)$. Based on 
the source type, we adopt the limb-darkening coefficient of $\Gamma_I=0.49$.

In the bottom panel of Figure~\ref{fig:one}, we present the residual of the data from 
the best-fit ``standard'' model. It is found that data around the main features of the 
light curve exhibit noticeable deviations. Considering that the time gap between the 
first and last peaks, $\sim 20$ days, is considerable, the residual can be ascribed to 
higher-order effects. We, therefore, conduct additional modeling considering the 
higher-order effects. In the ``parallax'' and ``orbit'' models, we separately consider 
the microlens-parallax effect and lens-orbital effect, respectively. In the 
``orbit+parallax'' model, we consider both effects.

Consideration of the microlens-parallax effect requires to include two additional 
parameters $\pi_{{\rm E},N}$ and $\pi_{{\rm E},E}$, which represent the two components 
of the microlens parallax vector $\pivec_{\rm E}$ projected onto the sky along the north 
and east equatorial coordinates, respectively.  To first-order approximation, the 
lens-orbital effect is describe by two parameters $ds/dt$ and $d\alpha/dt$, which 
represent the change rates of the projected binary separation $s$ and the source 
trajectory angle $\alpha$, respectively.  For the description of the full Keplerian 
orbital motion, one needs two more parameters $s_\parallel$ and $ds_\parallel/dt$, 
which are the line-of-sight separation between the binary components and its rate 
of change, respectively \citep{Skowron2011}. In our analysis, we consider the orbital 
effect with the two parameters $ds/dt$ and $d\alpha/dt$.

\begin{figure}[t]
\epsscale{1.0}
\plotone{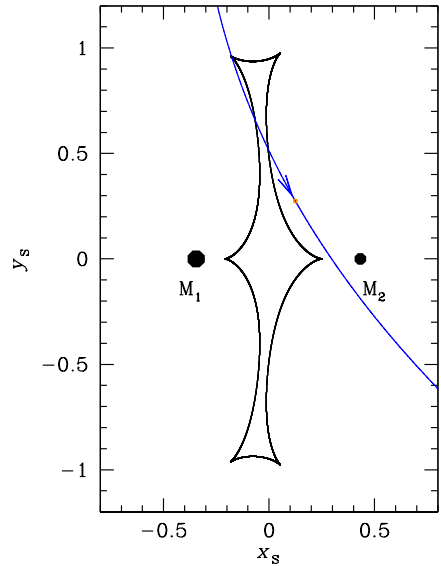}
\caption{\label{fig:two}
Geometry of the lens system. The curve with an arrow represents the source trajectory
with respect to the caustic which is the figure composed of concave curves.  The Two 
filled dots marked by $M_1$ and $M_2$ ($ < M_1$) are the locations of the binary lens 
components. Lengths are scaled to the Einstein radius corresponding to the total mass 
of the binary lens.  The size of the tiny circle at the tip of the arrow on the source 
trajectory represents the source size.  The model is based on the ``orbit+parallax'' 
model of which parameters are presented in Table~\ref{table:one}.  We note that the 
lens position and the resulting caustic vary in time due to the orbital motion of the 
lens and the presented positions are those at ${\rm HJD}=2457144$, which corresponds 
to the time of the second peak in the light curve presented in Fig.~\ref{fig:one}. 
}
\end{figure}

\begin{deluxetable}{crrrrr}
\tablecaption{Degenerate solutions\label{table:two}}
\tablewidth{0pt}
\tablehead{
\multicolumn{1}{c}{Solution}            &
\multicolumn{1}{c}{$\chi^2$}            &
\multicolumn{1}{c}{$\pi_{{\rm E},N}$}   &
\multicolumn{1}{c}{$\pi_{{\rm E},E}$}   &
\multicolumn{1}{c}{$ds/dt$}             &
\multicolumn{1}{c}{$d\alpha/dt$}          
}
\startdata
(1)   & 772.2    & -2.26  & -0.08  & -0.46  & -0.89 \\
(2)   & 772.5    & -3.32  &  0.00  & -0.34  & -0.39 \\
(3)   & 772.6    & -0.70  &  0.38  & -0.54  & -2.02 \\
(4)   & 776.0    &  0.05  &  0.15  &  0.15  &  0.17 
\enddata
\end{deluxetable}

In Table~\ref{table:one}, we summarize the results of modeling along with the 
$\chi^2$ values and the lensing parameters of the individual tested models.  
In the lower panels of Figure~\ref{fig:one}, we present the residuals of the 
models.   The model light curve corresponding to the best-fit model, i.e.\ 
orbit+parallax, is plotted over the data in Figure~\ref{table:one}.  In 
Figure~\ref{fig:two}, we also present the geometry of the lens system showing 
the source trajectory with respect to the caustic for the best-fit model.  
Figure~\ref{fig:two} shows that the event was produced by the passage of the 
source over a single large caustic formed by a resonant binary with a projected 
separation similar to the Einstein radius corresponding to the total mass of 
the binary, i.e.\ $s\sim 1$. The second peak was produced by the crossing of 
the source star over the caustic and first and third peaks were produced by 
the source star's approaches to the tips of the caustic.

From the results of analysis, it is found that the fit greatly improves when higher-order effects 
are considered. We find that the consideration of the microlens-parallax effect improves the fit 
by $\Delta\chi^2=839.7$ with respect to the standard model. The consideration of the lens-orbital 
effect results in a similar fit improvement with $\Delta\chi^2=835.1$. However, the further improvement 
of the fit by considering both effects is meager: $\Delta\chi^2=5.7$ from the parallax model and 
$\Delta\chi^2=10.3$ from the orbital model. The fact that the $\chi^2$ improvements by the tested 
models are similar one another regardless of the considered higher-order effects strongly indicates 
that the lens-parallax and lens-orbital effects are difficult to be distinguished despite the obvious 
influence of the higher-order effects on the lensing light curve.

\begin{figure}[t]
\epsscale{1.1}
\plotone{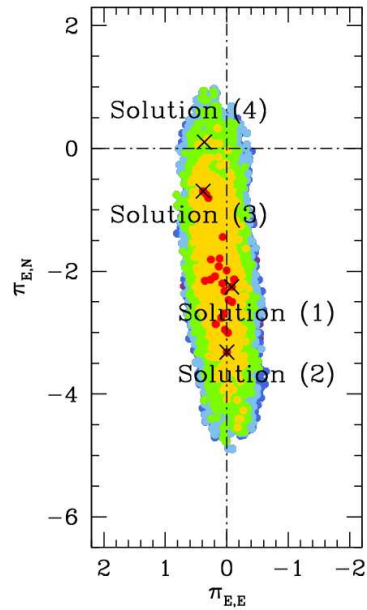}
\caption{\label{fig:three}
Distribution of the parallax parameters $\pi_{{\rm E},N}$ 
and $\pi_{{\rm E},E}$. obtained from the orbit+parallax modeling. The color coding 
represents points on the MCMC chain within 1 (red), 2 (yellow), 3 (green), 4 (cyan), 
and 5 (blue) $\sigma$ of the best fit. 
The microlens-parallax parameters, $\pi_{{\rm E},N}$ and $\pi_{{\rm E},E}$, and the 
lens-orbital parameters, $ds/dt$ and $d\alpha/dt$, corresponding to the 4 solutions 
designated by (1) through (4) on the plot are presented in Table~\ref{table:two}.
}
\end{figure}

The degeneracy between the microlens-parallax and lens-orbital effects is also shown 
in Figure~\ref{fig:three}, where we plot the distribution of the parallax parameters 
$\pi_{{\rm E},N}$ and $\pi_{{\rm E},E}$ obtained from the orbit+parallax modeling. 
The color coding represents points on the MCMC chain within 1$\sigma$ (red), 2$\sigma$ 
(yellow), 3$\sigma$ (green), 4$\sigma$ (cyan), and 5$\sigma$  (blue) of the best fit 
value. One finds that the observed light curve is explained with parallax parameters 
that are distributed in a wide range of the parameter space. On the 
$\pi_{{\rm E},N} - \pi_{{\rm E},E}$ distribution plot, we mark 4 different solutions 
for which the parallax parameters and orbital parameters are presented in Table~\ref{table:two} 
along with the corresponding $\chi^2$ values. For the solution with a large $\pi_{\rm E}$ 
value, e.g.\ `Solution (2)',  the deviation from the standard model is explained mostly 
by the parallax effect. On the other hand, the solution with a small $\pi_{\rm E}$ value, 
e.g.\ `Solution (4)', the deviation is mostly explained by the orbital effect. Despite 
the large difference between the parameters of the higher-order effects, $\chi^2$ 
differences among the solutions are very minor, indicating that the degeneracy between 
the microlens-parallax and lens-orbital effects are very severe.

\begin{figure*}[t]
\epsscale{0.95}
\plotone{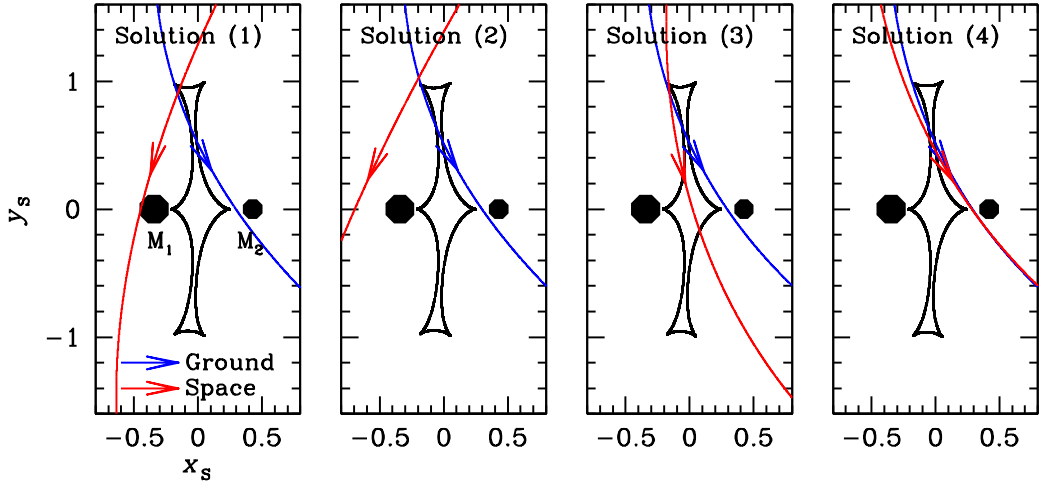}
\caption{\label{fig:four}
Lens system geometry corresponding to the 4 solutions presented in Table~\ref{table:two}. 
In each panel, the blue curve represents the source trajectory seen from the Earth and 
the red curve represents the source trajectory if the event were observed by the {\it Spitzer}
telescope. Other notations are same as those in Fig.~\ref{fig:two}.
}
\end{figure*}

\begin{figure}[t]
\epsscale{1.2}
\plotone{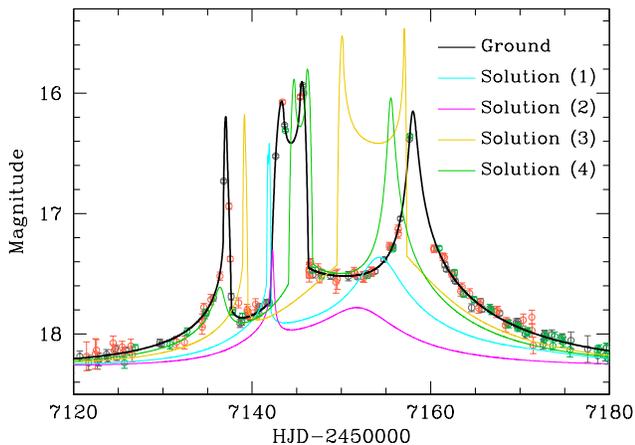}
\caption{\label{fig:five}
Expected light curves from space-based observations using the {\it Spitzer} telescope 
for the 4 solutions in Table~\ref{table:two}. Also presented is the best-fit 
model of the ground-based data. 
}
\end{figure}

\section{Resolving the Degeneracy}

In the previous section, we demonstrated that despite an obvious long-term deviation in a
microlensing light curve it could be difficult to identify the cause of the deviation 
because of the difficulty in distinguishing between the microlens-parallax and lens-orbital 
effects. In this section, we show that the degeneracy can be readily resolved with the additional 
data provided by space-based microlens parallax observations.

Resolving the degeneracy with the space-based data will be possible because the 
degenerate solutions with different values of the microlens parallax will result 
in different lensing light curves when an event is observed by the satellite. We 
illustrate this for OGLE-2015-BLG-0768 that was used to show the severity of the 
degeneracy between the microlens-parallax and lens-orbital effects in the data 
obtained from ground-based observations.  In Figure~\ref{fig:four}, we present 
the lens system geometry corresponding to the 4 solutions presented in Table~\ref{table:two}.  
In each panel, the blue curve represents the source trajectory seen from the Earth 
and the red curve represents the source trajectory that is expected if the event were 
observed by the {\it Spitzer} telescope.\footnote{We note that the event OGLE-2015-BLG-0768 
was not observed by {\it Spitzer} telescope because it was judged that the event could not 
be completed until the possible observation date of ${\rm HJD}'\sim 7180$ that was set by 
the sun-exclusion angle.} As expected, one finds that the ground-based source trajectories 
of the individual solutions show similar paths with respect to the caustics, resulting in 
similar ground-based light curves. In contrast, the space-based trajectories have 
dramatically different paths from one another due to the differences in the values of 
the microlens parallax.

In Figure~\ref{fig:five}, we present the space-based light curves corresponding to the 
source trajectories that are marked in the individual panels of Figure~\ref{fig:four}.
In order to show the level of difference between the ground-based and space-based light 
curves, we also present the ground-based data and the best-fit model.  From the comparison 
of the light curves, it is found that despite the similarity in the ground-based light 
curves, the resulting space-based light curves of the individual solutions show dramatically 
different shapes.  This indicates that the microlens parallax can be uniquely determined 
from the difference between the ground-based and space-based light curves and thus the 
severe degeneracy in the interpretation of the ground-based data can be readily resolved 
with the additional data provided by the space-based microlensing observations.  Once 
the microlens parallax is determined, one can further constrain the orbital parameters 
of the lens by analyzing remaining residuals from the parallax model.  Therefore, data 
from space-based microlens-parallax observations are important not only in accurately 
determining the basic lens parameters of the mass and distance but also in characterizing 
the orbital parameters of the lens.

The possibility of characterizing the orbital parameters of a binary lens was recently 
pointed out by \citet{Han2016}, where they presented the analysis of the combined data 
from the ground-based and space-based {\it Spitzer} observations of the binary-lens 
event OGLE-2015-BLG-0479.  Thanks to the uniquely determined microlens parallax with 
the space-based data, they were able to constrain the complete orbital parameters of 
the lens, although the uncertainties of the estimated orbital parameters are rather 
big due to the partial coverage of the event by the {\it Spitzer} data combined with 
the sparse coverage and modest photometry quality of the ground-based data.  The 
precision will be improved with the expansion of both the ground-based and space-based 
surveys.

\section{Conclusion}

We demonstrated 
that interpretation of long-term deviations in lensing light curves could be difficult due to the severe 
degeneracy between the microlens-parallax and lens-orbital effects even for the case of obvious deviations. 
We also showed that the degeneracy could be readily resolved with the additional data from
space-based microlens parallax observations. Being able to unambiguously determine the microlens parallax, 
space-based microlens observations will enable to determine the physical parameters of the lens with 
an increased accuracy. Furthermore, space-based data will make it possible to constrain the orbital 
parameters of the lens with an unprecedented precision.

\acknowledgments
Work by C.~Han was supported by the Creative Research Initiative Program (2009-0081561) of 
the National Research Foundation of Korea.  
The OGLE project has received funding from the National Science Centre, Poland, grant 
MAESTRO 2014/14/A/ST9/00121 to AU.  OGLE Team thanks Profs.\ M.~Kubiak and G.~Pietrzy{\'n}ski, 
former members of the OGLE team, for their contribution to the collection of the OGLE 
photometric data over the past years.
Work by AG was supported by JPL grant 1500811.
We acknowledge the high-speed internet service (KREONET)
provided by Korea Institute of Science and Technology Information (KISTI).
The KMTNet telescopes are operated by the Korea Astronomy and Space Science Institute (KASI).

\end{document}